\documentclass[12pt]{article}
\usepackage{amssymb,amsmath}

\newcommand{\calpsi}{\mbox{$\mathsf{\Psi}_{\displaystyle{\hspace{-1.67ex}\mathsf{J}}}$}}

\begin{document}

\title{{\bf The Orbifolds of Permutation-Type as\\
Physical String Systems\\
 at Multiples of $\mathbf{c=26}$\\
II.  The Twisted BRST Systems of $\mathbf{\hat c = 52}$ Matter}}
\author{M.~B. Halpern\footnote{halpern@physics.berkeley.edu}\\
Department of Physics, 
University of California, \\
and Theoretical Physics Group, \\
Lawrence Berkeley National Laboratory\\
University of California, 
Berkeley, California 94720, USA}

\maketitle

\begin{abstract}
This is the second in a series of papers which consider the orbifolds
of permutation-type as candidates for new physical string systems at higher central charge.
In the first paper, I worked out the extended actions of the twisted sectors
of those orbifolds -- which exhibit new permutation-twisted world-sheet gravities
and correspondingly extended diffeomorphism groups.  In this paper I begin the study
of these systems as operator string theories, limiting the discussion
 for simplicity to
the strings with ${\hat c} = 52$ matter (which are those governed by ${\mathbb Z}_2$-twisted
permutation gravity).  In particular, I present here a construction of the twisted
reparametrization ghosts and {\em new twisted BRST systems} of all ${\hat c} = 52$ strings.
The twisted BRST systems also imply new {\em extended physical state 
conditions}, whose analysis
for individual ${\hat c} = 52$ strings is deferred to the next paper of the series.
\end{abstract}

\clearpage
\tableofcontents

\clearpage
\section{Introduction}
\label{sec1}

Opening a third, more phenomenonological chapter in the orbifold 
program [1-11,12-15], I have speculated in Ref. [16] that the orbifolds of 
permutation-type may describe physical string systems at higher 
critical central charge. This includes the values
\renewcommand{\theequation}{\thesection.\arabic{equation}}
\setcounter{equation}{0}
\begin{equation}
\label{eq1.1}
{\hat c} = 26K,\,\,\, K = 2,3,\dots
\end{equation}
for the bosonic cases and
\begin{equation}
\label{eq1.2}
{\hat c} = (10K),\quad\,\,\, ({\hat c},{\hat {\bar c}}) = (26K,10K)
\end{equation}
for superstring orbifolds of permutation-type. There is a surprisingly 
large variety of orbifolds of permutation-type, including the 
permutation orbifolds themselves [1-6,11,16], the orientation orbifolds 
[12,13,15, 16], the 
open-string permutation orbifolds [14] and their $T$-duals [15,16], the 
generalized permutation orbifolds [15,16] and others. A short review of these 
varieties is included in Ref. [16].

The technical observation which underlies this conjecture is as 
follows: The extended world-sheet actions of the twisted sectors of 
the bosonic orbifolds exhibit a large class of new extended (twisted) world-sheet
gravities -- called the {\it permutation gravities} [16] -- which are
accompanied by correspondingly-extended diffeomorphism invariances. 
Like the sectors of the permutation orbifolds, the permutation 
gravities are classified by the conjugacy classes of the permutation 
groups $H(\mbox{perm})$ -- where the trivial class of $H(\mbox{perm})$ 
corresponds to (decoupled copies of) ordinary Polyakov gravity in the untwisted 
sectors of the orbifolds. Indeed, the permutation gravities can be 
understood \footnote {See in particular Subsec. 3.5 of Ref. [16].} as nothing
 but the standard orbifold map
(the principle of local isomorphisms) of 
ordinary Polyakov gravity into the twisted sectors. The explicit forms of the 
corresponding permutation supergravities of the superstring orbifolds 
have not yet been worked out.

The permutation gravities play the following role in the conjecture: 
The covariant formulation of the twisted sectors of the orbifolds
 of permutation-type
 exhibit an increased 
number of negative-norm states (time-like currents) in correspondence 
with their higher central charges. This parallels the situation in the 
untwisted sectors, where the negative-norm states of the untwisted 
string copies are removed by the ordinary Polyakov gravity of each 
copy. Since the permutation gravities are the maps of the untwisted 
gravities into the twisted sectors of the orbifold, one may expect the 
extended diffeomorphism invariances of the new gravities to similarly 
remove the negative-norm states of each twisted sector. The size of 
the new twisted diffeomorphism invariances lends strong support to 
this speculation.

The classical intuition of Ref. [16] must of course be verified both at 
the operator level and the interacting level, work which I begin in 
the present paper. In particular, I construct here the twisted 
reparametrization ghosts and {\it new twisted BRST systems} for all the 
twisted open- and closed-string sectors which are governed by the 
simple case of $\mathbb{Z}_{2}$-twisted permutation gravity, i.e.
all twisted strings with matter central charge  ${\hat c} = 52$.

The new BRST systems then directly imply new {\it extended physical 
state conditions} for each $\hat c=52$ string, which are in fact the 
operator realizations of the classical extended Virasoro constraints 
of Ref. [16]. I emphasize that the number of these conditions at $\hat c=52$
is doubled relative to the 
ordinary physical state condition of $c=26$ strings, so the counting 
at the operator level still appears adequate to eliminate the doubled 
set of negative-norm states at $\hat c=52$. For brevity however, 
analysis of the extended physical state conditions and spectrum of 
particular $\hat c =52$ strings is deferred to the next paper of 
this series.

I should also emphasize that the construction of this paper is largely self-contained,
using only standard operator-product techniques in the orbifold program. 
Moreover, given the operator-product formulation of untwisted BRST in 
Sec. 2, the computation given in the succeeding sections can be straightforwardly
 generalized to obtain the twisted
BRST systems of all the orbifolds of permutation-type. Alternately, these systems
can be constructed $\grave a$ la Faddeev-Popov from the extended actions of 
Ref. [16], but I will not follow this line here.

\vspace{.4in}
\setcounter{section}{1}
\section{Operator-Product Form of Untwisted BRST}
\label{sec2}

In order to apply the techniques of the orbifold program, one first 
needs the ordinary BRST system [17-19] in operator-product form. The simplest 
approach to this is the picture of Freericks and Halpern [20], where 
the (left-mover) reparametrization ghosts [21] are understood as a conformal 
deformation \footnote {General conformal deformations [20, 22] come in two 
varieties, c-changing [24,25] and c-fixed -- the latter now often called 
spectral flow [25, 26]. The so-called  ${\mathfrak sl}(2)$-preserving 
conformal deformations are those particular linear combinations of the two varieties 
which preserve the  ${\mathfrak sl}(2)$-invariant ground state of the 
undeformed theory (see Eq. (2.1i)).}
of ordinary complex half-integer moded Bardakci-Halpern (BH) fermions [26]:
\renewcommand{\theequation}{\thesection.\arabic{subsection}\alph{equation}}
\setcounter{subsection}{1}
\setcounter{equation}{0}
\begin{equation}
\label{eq2.1a}
\psi(z) = \sum_{p \in {\mathbb Z} - \frac {1}{2}} \psi(p)z^{-p-\frac 
{1}{2}},\quad {\bar \psi}(z) = \sum_{p \in {\mathbb Z} - \frac {1}{2}} {\bar \psi}(p)z^{-p-\frac {1}{2}}
\end{equation}
\begin{equation}
\label{eq2.1b}
[\psi(p),{\bar \psi}(q)]_+ = \delta_{p+q,0} 
\end{equation}
\begin{equation}
\label{eq2.1c}
\psi(p > 0) |0\rangle\, =\, {\bar \psi}(p > 0) |0\rangle \, = 0 
\end{equation}
\begin{equation}
\label{eq2.1d}
J(z) \equiv \,\,\, :\!{\bar\psi}(z) \psi(z)\!: \,\,\,=\,\, \sum_{m \in {\mathbb Z}} J(m)z^{-m-1} 
\end{equation}
\begin{equation}
\label{2.1e}
T^{BH}(z) \equiv \,\,\, -\tfrac {1}{2} :\!{\bar 
\psi}(z)\stackrel{\leftrightarrow}{\partial}\psi(z)\!:\,\,\, =\,\, \sum_{m} 
L^{BH}(m)z^{-m-2}
\end{equation}
\begin{eqnarray}
\label{eq2.1f}
T^G(z) &\equiv & T^{BH}(z) + \tfrac {3}{2} \partial J(z) \\
\label{eq2.1g}
&= &:\!{\bar \psi}(z)\partial\psi(z) + 2\partial{\bar \psi}(z)\psi(z)\!: \\
\label{eq2.1h}
&= &\sum_m L^G(m)z^{-m-2}
\end{eqnarray}
\begin{equation}
\label{eq2.1i}
L^{G}(m \ge -1)|0\rangle = L^{BH}(m \ge -1)|0\rangle = 0. 
\end{equation}
Here $|0\rangle$ is the BH vacuum and $T^{BH}$ is the BH stress tensor 
with central charge 26. The ghost stress tensor $T^G$ is a  conformal deformation by $\partial J$ of the BH 
stress tensor. In the language of Ref. [20], $\partial J$ is an example of 
a c-changing deformation which, according to Eq. (2.1i), is also an ${\mathfrak sl}(2)$-preserving deformation.
For the later transition to the orbifolds, it is important that all 
relations be written in terms of operator-product normal ordering 
 $:\,\cdot\,:$
(subtract singular terms) -- which in this case is equal to BH-mode normal ordering
\setcounter{subsection}{2}
\setcounter{equation}{0}
\begin{equation}
\label{eq2.2a} 
:\!A(p)B(q)\!:_M \,\,\equiv -\theta(p>0)B(q)A(q)\, +\, \theta(p < 0)A(p)B(q) 
\end{equation}
\begin{equation}
\label{eq2.2b} 
:\!A(z)B(\omega)\!:\,\,\, =\,\,\, :\!A(z)B(\omega)\!:_M 
\end{equation}
where $A$ and $B$ can be $\psi$ or$\bar\psi$.

 The operator product description of the ghost system is:
\setcounter{subsection}{3}
\setcounter{equation}{0}
\begin{equation}
\label{eq2.3a}
\psi(z){\bar \psi}(\omega) =\,\, \frac {1}{z-\omega}\,\,\,\, + :\!\psi(z){\bar 
\psi}(\omega)\!: 
\end{equation}
\begin{equation}
\label{eq2.3b}
T^G(z){\bar \psi}(\omega) = (-\frac {1}{(z-\omega)^2}\,
 + \frac {1}{z-\omega} \partial_{\omega}){\bar \psi}(\omega)\,\,\, + 
 :\!T^G(z){\bar \psi}(\omega)\!: 
\end{equation}
\begin{equation}
\label{eq2.3c}
T^G(z)\psi(\omega) = ( \frac {2}{(z-\omega)^2}\,\, + \frac {1}{z-\omega} \partial_{\omega}) \psi(\omega)
\,\,\, + :\!T^G(z)\psi(\omega)\!: 
\end{equation}
\begin{eqnarray}
\label{eq2.3d}
T^G(z)T^G(\omega) &=& -\frac {26/2}{(z-\omega)^4} + ( \frac {2}{(z-\omega)^2} +
 \frac {1}{z-\omega} \partial_{\omega}) T^G(\omega)\,\, + \nonumber \\ 
 & & \hspace{.1in} +:\!T^G(z)T^G(\omega)\!:. 
\end{eqnarray}
These relations follow easily from the Wick expansion of the BH 
fermions, and show that $T^{G}$ has central charge -26. 
Following the discussion of Ref. [20], the usual ghost modes $\{c\}$ and 
$\{b\}$ are then identified by relabelling (spectral flow) of the BH modes
\setcounter{subsection}{4}
\setcounter{equation}{0}
\begin{equation}
\label{eq2.4a}
{\bar \psi}(z) = \sum_m c(m)z^{-m+1},\quad \psi(z) = \sum_m b(m)z^{-m-2} 
\end{equation}
\begin{equation}
\label{eq2.4b}
c(m) \equiv {\bar \psi}( m - \tfrac {3}{2}),\quad b(m) \equiv \psi( m + 
\tfrac {3}{2}),\quad m \in {\mathbb Z} 
\end{equation}
\begin{equation}
\label{eq2.4c}
[c(m),b(n)]_+ = \delta_{m+n,0}
\end{equation}
\begin{equation}
\label{eq2.4d}
c(m \ge 2)|0\rangle\,\,\, =\,\,\, b(m \ge -1)|0\rangle\,\, =\,\, 0
\end{equation}
\begin{equation}
\label{eq2.4e}
[L^G(m),c(n)] = -(2m+n)c(m+n)
\end{equation}
\begin{equation}
\label{eq2.4f}
[L^G(m),b(n)] = (m-n)b(m+n)
\end{equation}
where $|0\rangle$ is still the BH vacuum. The commutators in Eqs. (2.4e) and 
(2.4f) show that the $\{c,b\}$ modes (and not the BH modes) have conformal 
weights $\Delta = -1,2$ respectively under $T^{G}$. 

To go further in this direction, one can rewrite the composite 
operators in terms of the ghost modes, introducing the standard 
Ramond-type mode ordering
\setcounter{subsection}{5}
\setcounter{equation}{0}
\begin{eqnarray}
\label{eq2.5a}
:\!c(m)b(n)\!:_R &\equiv &-\theta(m > 0)b(n)c(m) + \tfrac {1}{2} \delta_{m,0}[c(0),b(0)] \nonumber \\
& &\hspace{.2in}+ \theta(m<0)c(m)b(n) 
\end{eqnarray}
\begin{equation}
\label{eq2.5b}
J^G(z) \equiv\,\, :\!{\bar \psi}(z)\psi(z)\!:_R\,\, =\,\, J(z) - \tfrac 
{3}{2z}\,\, =\,\, \sum_m J^G(m)z^{-m-1}
\end{equation}
\begin{equation}
\label{eq2.5c}
J^G(m=0) = \tfrac {1}{2} [c(0),b(0)] + \sum_{n=1}^{\infty} (c(-n)b(n)-b(-n)c(n))
\end{equation}
which is needed to define the standard ghost current $J^G$ in Eq. 
(2.5b). There is no need to pursue this familiar line here however
because, as noted above, local operators (not modes) and operator-product normal 
ordering (not Ramond ordering) are required to apply the 
principle of local isomorphisms in Section 4.

Returning then to the local formulation, I further introduce the (left-mover) 
matter stress tensor $T$ at $c =26$ of the critical closed bosonic 
string, and the BRST current $J_{B}$:
\setcounter{subsection}{6}
\setcounter{equation}{0}
\begin{equation}
\label{eq2.6a}
T(z)T(\omega) = \frac {13}{(z-\omega)^4} + ( \frac {2}{(z-\omega)^2} + \frac {1}{z-\omega} \partial_{\omega})T(\omega)
\,\, + :\!T(z)T(\omega)\!:
\end{equation}
\begin{equation}
\label{eq2.6b}
J^B(z) \equiv \,\,{\bar \psi}(z)T(z) + \tfrac {1}{2} :\!{\bar 
\psi}(z)T^G(z)\!:\,\,\, = \sum_m J^B(m)z^{-m-1}.
\end{equation}
After some algebra with the Wick expansion for BH fermions, one finds the desired 
operator products
\setcounter{subsection}{7}
\setcounter{equation}{0}
\begin{equation}
\label{eq2.7a}
J^B(z)\psi(\omega) = \frac {J(\omega)}{(z-\omega)^2} + \frac 
{T^t(\omega)}{z-\omega}\,\, + :\!J^B(z)\psi(\omega)\!:
\end{equation}
\begin{equation}
\label{eq2.7b}
T^t(z) \equiv T(z) + T^G(z) = \sum_m L^t(m)z^{-m-2}
\end{equation}
\begin{eqnarray}
\label{eq2.7c}
J^B(z)J^B(\omega) &= &\frac {10}{(z-\omega)^3} \partial_{\omega} {\bar \psi}(\omega){\bar \psi}(\omega)
 + \frac {5}{(z-\omega)^2} \partial_{\omega}^2{\bar \psi}(\omega){\bar \psi}(\omega) \nonumber \\
& &+ \frac {3}{2(z-\omega)} \partial_{\omega} (\partial_{\omega}^2{\bar 
\psi}(\omega){\bar \psi}(\omega))\,\, + :\!J^B(z)J^B(\omega)\!: 
\end{eqnarray}
where $T^t$ in Eq. (2.7b) is the total stress tensor with zero central 
charge.

This (and a right-mover copy of the system) completes the operator- 
product description needed below but, because Eq. (2.7c) is new, I also 
give selected parts of the mode algebra corresponding to these results:
\setcounter{subsection}{8}
\setcounter{equation}{0}
\begin{equation}
\label{eq2.8a}
[J^B(m),b(n)]_+ =\,\, mJ(m+n) \,+\, L^t(m+n)
\end{equation}
\begin{eqnarray}
\label{eq2.8b}
[J^B(m),J^B(n)]_+ &= &\tfrac {1}{2} (10mn - 3(m+n)(m+n-1))\times \nonumber \\
& &\quad \times\sum_{p \in {\mathbb Z}} pc(p)c(m+n+p)
\end{eqnarray}
\begin{equation}
\label{eq2.8c}
Q \equiv J^B(m=0)
\end{equation}
\begin{equation}
\label{eq2.8d}
[Q,b(m)]_+ = L^t(m),\quad [Q,L^{t}(m)] =0.
\end{equation}
\begin{equation}
\label{eq2.8e}
 Q^2 = 0
\end{equation}
Besides the standard BRST charge $Q$, I note in passing that 
other nilpotent operators such as $(J^B(3))^2 = 0$ are implied by the 
anticommutator (2.8b).

\setcounter{section}{2}
\section{Automorphisms and Eigenfields}
\label{sec3}

In this and the following section, I will apply standard operator 
techniques in the orbifold program to construct the twisted BRST 
systems of all $\hat c=52$ matter. These systems comprise all the 
twisted sectors of the orbifolds of permutation-type which are 
governed by $\mathbb{Z}_{2}$-twisted permutation gravity [16], or 
equivalently, those 
which are governed by an order-two orbifold Virasoro 
algebra at $\hat c=52$ [1,27,9,12,16]. The list of these sectors includes the twisted 
open-string sectors of the orientation orbifolds and the twisted 
closed-string sectors of the generalized $\mathbb{Z}_{2}$-permutation 
orbifolds
\renewcommand{\theequation}{\thesection.\arabic{subsection}\alph{equation}}
\setcounter{subsection}{1}
\setcounter{equation}{0}
\begin{eqnarray}
\label{eq3.1a}
\frac {U(1)^{26}}{H_-},\,\,\,\,\,\, H_-\hspace{-.1in} &= &{\mathbb Z}_2(\mbox{w.s.}) \times H \\
\label{eq3.1b}
\frac {U(1)^{26}\times U(1)^{26}}{H_+},\,\,\,\,\,\, H_+\hspace{-.1in} &= &{\mathbb Z}_2(\mbox{perm}) \times H' 
\end{eqnarray}
as well as the generalized open-string $\mathbb{Z}_{2}$-permutation 
orbifolds and their $T$-duals \footnote {The generalized open-string 
permutation orbifolds and their T-duals [15] are constructed from the 
left-mover data of the generalized permutation orbifolds 
$U(1)^{26K}/(H(\mbox{perm})\times H')$). The open-string orientation- 
orbifold sectors  are among the $T$-duals of the generalized 
open-string $\mathbb{Z}_{2}$-permutation orbifolds.}. In Eq. (3.1), $H_{\mp}$ 
are automorphism groups generated by $\tau_{\mp} \times \omega$, $\omega 
\in H$ or $H'$, where ${\tau}_{-}$ is the element of 
${\mathbb Z}_{2}(\mbox{w.s.})$ which exchanges the left- and 
right-movers of the untwisted closed string $U(1)^{26}$ and 
${\tau}_{+}$ is the element of $\mathbb{Z}_{2}(\mbox{perm})$ which exchanges the two copies of the closed-string. In both 
cases, the extra automorphisms $\omega$ act uniformly on the left- and right-movers of each closed 
string.

Both orbifold types in Eq. (3.1) can have many twisted sectors, 
labelled by the conjugacy classes of $H_{\mp}$, and half the twisted 
sectors of the orientation orbifolds (corresponding to the 
orientation-reversing automorphisms of $H_{-}$) are twisted open 
string CFT's at $\hat c=52$. I will have more to say about the 
relation of orientation orbifolds to orientifolds [28] in succeeding 
papers of the series.  The extended Polyakov actions (with ${\mathbb 
Z}_2$-twisted permutation gravity) of all these sectors are given in 
Eqs. (2.52) and (4.6) of Ref. [16].

The first step in the orbifold program is to specify the action of the 
automorphisms on the untwisted fields. The action on the matter fields has been 
studied elsewhere [9,12,16], so I focus here on the fields of the BRST 
system which (like the extended stress tensors and $\mathbb{Z}_{2}$-twisted permutation gravity 
itself) see only the  ${\mathbb Z}_2$'s (${\mathbb Z}_2(\mbox{w.s.})$ or ${\mathbb 
Z}_2(\mbox{perm})$) of the examples in Eq. (3.1). Following the 
development of Refs. [12,13], we may summarize the action of the 
non-trivial element of each ${\mathbb Z}_2$ in the following {\it unified, 
two-component} notation:
\renewcommand{\theequation}{\thesection.\arabic{subsection}\alph{equation}}
\setcounter{subsection}{2}
\setcounter{equation}{0}
\begin{equation}
\label{eq3.2a}
T_I^G \equiv\,\,\, :\!{\bar \psi}_I\partial\psi_I + 2\partial{\bar 
\psi}_I\psi_I\!:,\quad\, I = 0,1
\end{equation}
\begin{equation}
\label{eq3.2b}
J_I =\,\,\, :\!{\bar \psi}_I\psi_I\!:,\quad\,\, J_I^B = {\bar 
\psi}_IT_I + \tfrac {1}{2} :\!{\bar \psi}_IT_I^G\!: 
\end{equation}
\begin{equation}
\label{eq3.2c}
A_I(z)' = (\tau_1)_I{}^J A_J(z),\quad\,\, \tau_1 = \begin{pmatrix} 0 & 1 \\ 1 & 0 \end{pmatrix}.
\end{equation}
Here all the operators are functions of complex $z$ and ${A_I}'$ in 
(3.2c) is the automorphic response of any operator in the system.

I want to emphasize that, with appropriate identifications, the unified action (3.2) describes
 the BRST systems 
of {\it all} untwisted sectors which can lead to twisted sectors at $\hat c=52$. In the case of the 
generalized  ${\mathbb Z}_2$-permutation orbifolds (3.1b), as well as 
the generalized open-string $\mathbb{Z}_{2}$-permutation orbifolds 
[15], the index $I$ labels the two left-mover copies in $U(1)^{26} 
\times U(1)^{26}$ (or the two right-mover copies with $z \to {\bar z}$ 
for the permutation orbifolds). For the orientation orbifolds, $0$ 
and $1$ label the left- and right-movers respectively of the single 
closed string $U(1)^{26}$, where the right movers are relabelled by the ${\bar z} \to 
z$ trick of Refs. [12,13]
\renewcommand{\theequation}{\thesection.\arabic{equation}}
\setcounter{equation}{2}
\begin{equation}
\label{eq3.3}
A_0(z) \equiv A(z),\,\,\, A_1(z) \equiv {\bar A}({\bar z})|_{{\bar z} \to z}
\end{equation}
which preserves the desired exchange of the left- and right-mover 
modes.

Then one easily finds the operator products of the two-component 
fields, e.g.
\renewcommand{\theequation}{\thesection.\arabic{subsection}\alph{equation}}
\setcounter{subsection}{4}
\setcounter{equation}{0}
\begin{equation}
\label{eq3.4a}
{\bar \psi}_I(z)\psi_J(\omega) = \frac {\delta_{IJ}}{z-\omega}\,\, + 
:\!{\bar \psi}_I(z)\psi_J(\omega)\!:
\end{equation}
\begin{eqnarray}
\label{eq3.4b}
T_I^G(z)T_J^G(\omega) &=& \delta_{IJ}( \frac {-13}{(z-\omega)^4} + ( \frac {2}{(z-\omega)^2} +
 \frac {1}{z-\omega} \partial_{\omega}) T_I^G(\omega)) + \nonumber \\
& &\hspace{.1in} + :\!T_I^G(z)T_J^G(\omega)\!:
\end{eqnarray}
where $\delta_{IJ}$ is Kronecker delta. All the other two-component 
operator products have the same semi-simple structure, and it is easily 
checked that (3.2c) is an automorphism of the full two-component system.

The next step in the orbifold program is the definition of {\it eigenfields}, 
which diagonalize the automorphic response \footnote{Here I am using 
the notation $\bar u=0,1$ of the orientation orbifolds, while the 
standard notation for the $\mathbb{Z}_{2}$-permutation orbifolds is 
$\bar u \rightarrow \bar{\hat\jmath}=0,1$.}
\renewcommand{\theequation}{\thesection.\arabic{subsection}\alph{equation}}
\setcounter{subsection}{5}
\setcounter{equation}{0}
\begin{equation}
\label{eq3.5a}
U = U^{\dag} = \tfrac {1}{\sqrt{2}} \begin{pmatrix} 1 & 1 \\ 1 & -1 \end{pmatrix}
\end{equation}
\begin{equation}
\label{eq3.5b}
{\bar \calpsi}_u \equiv U_u{}^I{\bar \psi}_I,\quad \calpsi_u \equiv 
U_u{}^I\psi_I,\quad {\bar u} = 0,1
\end{equation}
\begin{equation}
\label{eq3.5c}
{\mathcal J}_u \equiv \sqrt{2} U_u{}^IJ_I = \sum_{v=0}^1 :\!{\bar 
\calpsi}_{u-v}\calpsi_v\!:
\end{equation}
\begin{equation}
\label{eq3.5d}
\theta_u^G \equiv \sqrt{2} U_u{}^I T_I^G = \sum_v :\!{\bar \calpsi}_{u-v} \partial\calpsi_v
 + 2\partial {\bar \calpsi}_{u-v}\calpsi_v\!:
\end{equation}
\begin{equation}
\label{eq3.5e}
\theta_u^t \equiv \sqrt{2} U_u{}^I (T_I + T_I^G) = \theta_u + \theta_u^G
\end{equation}
\begin{equation}
\label{eq3.5f}
{\mathcal J}_u^B \equiv 2U_u{}^I J_I^B = \sum_v ({\bar 
\calpsi}_{u-v}\theta_v + :\!{\bar \calpsi}_{u-v}\theta_v^G\!:)
\end{equation}
\begin{equation}
\label{eq3.5g}
{\mathcal A}_u(z)' = (-1)^u{\mathcal A}_u(z)
\end{equation}
where ${\mathcal A}_{u}$ in Eq. (3.5g) can be any eigenfield in the 
system. It is then straightforward to compute the operator products in 
the eigenfield basis, for example:
\renewcommand{\theequation}{\thesection.\arabic{subsection}\alph{equation}}
\setcounter{subsection}{6}
\setcounter{equation}{0}
\begin{equation}
\label{eq3.6a}
{\bar \calpsi}_u(z)\calpsi_v(\omega) = \frac {\delta_{u+v,0\!\!\!\!\mod 
2}}{z-\omega}\,\,\,\,\, + :\!{\bar \calpsi}_u(z)\calpsi_v(\omega)\!:
\end{equation}
\begin{eqnarray}
\label{eq3.6b}
\theta_u^G(z)\theta_v^G(\omega) &= &-\frac {52/2}{(z-\omega)^4} 
\delta_{u+v,0\!\!\!\!\mod 2}+\nonumber \\
 & &\hspace{-.4in}+ ( \frac {2}{(z-\omega)^2} + \frac {1}{z-\omega} \partial_{\omega})\theta_{u+v}^G(\omega)
 \,\,\,\,\, + :\!\theta_u^G(z)\theta_v^G(\omega)\!:.
\end{eqnarray}
I omit the rest of the eigenfield operator products for brevity, but 
the full set can easily be read from the result (4.2) of the following 
section. With Ref. [12], I remark 
in particular on the central charge $c=-52$ of the  ghost eigen-stress 
tensors $\{\theta_u^G$, $u = 0,1\}$ in Eq. (3.6b), which follows because 
$\theta_0^G = T_0^G + T_1^G$ and the central charges of the copies $T_{0,1}^{G}$ 
are additive. Similarly, the central charge of the matter eigen-stress 
tensors $\{\theta_{u}\}$ is $52$, and the anomaly cancels for the total 
eigen-stress tensors
$\{\theta_{u}^{t}\}$.

\setcounter{section}{3}
\section{The New Twisted BRST Systems}
\label{sec4}
The next step in the program is the transition to the orbifold, using 
the {\it principle of local isomorphisms}  [1,3,5,6,9,11,12]. This principle maps the 
eigenfields $\{\mathcal A\}$ of each conjugacy class of the automorphism 
group to the corresponding twisted fields $\{\hat A\}$ of each twisted sector,
 taking the phases of the automorphic response of the eigenfields as 
the monodromies of the twisted fields \footnote{The fields $A,\mathcal A, \hat A$ and the familiar fields 
$\hat{\mathcal A}$ with twisted boundary conditions form  commuting 
diagrams (see Refs. [3,6,11]), where the twisted fields $\hat A$
are the monodromy decomposition of $\hat{\mathcal A}$.}. To complete the 
transition, the
principle also specifies that the operator products
of the twisted fields are isomorphic to those of the eigenfield basis, 
and hence that operator-product normal-ordered products in the 
eigenfield basis map to 
operator-product normal-ordered products in the twisted sectors of the 
orbifold.  Correspondingly, central charges do not 
change under orbifoldization from the eigenfield basis.

In the present application, we then obtain the operators of the 
new twisted BRST systems
\renewcommand{\theequation}{\thesection.\arabic{subsection}\alph{equation}}
\setcounter{subsection}{1}
\setcounter{equation}{0}
\begin{equation}
\label{eq4.1a}
{\hat J}_u = \sum_v :\!{\hat {\bar \psi}}_{u-v}{\hat 
\psi}_v\!:,\quad\,\, {\hat \theta}_u^t = {\hat \theta}_u + {\hat \theta}_u^G
\end{equation}
\begin{equation}
\label{eq4.1b}
{\hat \theta}_u^G = -\tfrac {1}{2} \sum_v :\!{\hat {\bar \psi}}_{u-v} \stackrel{\leftrightarrow}{\partial}
 {\hat \psi}_v\!: +\,\, \tfrac {3}{2} {\hat J}_u
\end{equation}
\begin{equation}
\label{eq4.1c}
{\hat J}_u^B = \sum_v ({\hat {\bar \psi}}_{u-v} {\hat \theta}_v + :\!{\hat 
{\bar \psi}}_{u-v}{\hat \theta}_v^G\!:)
\end{equation}
\begin{equation}
\label{eq4.1d}
{\hat A}_u(ze^{2\pi i}) = (-1)^u {\hat A}_u(z),\quad {\bar u} = 0,1
\end{equation}
where ${\hat A}_u$ in (4.1d) can be any operator in the system.
Moreover, we obtain 
the twisted operator products of each sector \footnote {Supplementing 
the examples in Eq. (3.6), the complete set of eigenfield operator 
products can be read off from the result Eq. (4.2) by the substitution
 ${\hat A}_u \to  {\mathcal A}_u$.}:
\renewcommand{\theequation}{\thesection.\arabic{subsection}\alph{equation}}
\setcounter{subsection}{2}
\setcounter{equation}{0}
\begin{equation}
\label{eq4.2a}
{\hat {\bar \psi}}_u(z){\hat \psi}_v(\omega) = \frac 
{\delta_{u+v,0\!\!\!\!\mod 2}}{z-\omega}\,\,\,\, + :\!{\hat {\bar 
\psi}}_u(z){\hat \psi}_v(\omega)\!:\quad \bar u,\bar v\in\{0,1\}
\end{equation}
\begin{eqnarray}
\label{eq4.2b}
{\hat \theta}_u(z){\hat \theta}_v(\omega) &= &\frac {52/2}{(z-\omega)^4} \delta_{u+v,0\!\!\!\!\mod 2} 
+ ( \frac {2}{(z-\omega)^2} + \frac {1}{z-\omega} \partial_{\omega}) {\hat \theta}_{u+v}(\omega) +\nonumber \\
& &\hspace{.2in}+ :\!{\hat \theta}_u(z){\hat \theta}_v(\omega)\!:
\end{eqnarray}
\begin{eqnarray}
\label{eq4.2c}
{\hat \theta}_u^G(z){\hat \theta}_v^G(\omega) &= &-\frac {52/2}{(z-\omega)^4} \delta_{u+v,0\!\!\!\!\mod 2}
 + ( \frac {2}{(z-\omega)^2} + \frac {1}{z-\omega} \partial_{\omega}) 
 {\hat \theta}_{u+v}^G(\omega)+ \nonumber \\
& &\hspace{.2in}+  :\!{\hat \theta}_u^G(z){\hat \theta}_v^G(\omega)\!:
\end{eqnarray}
\begin{equation}
\label{eq4.2d}
{\hat \theta}_u^t(z){\hat \theta}_v^t(\omega) = ( \frac {2}{(z-\omega)^2} 
+ \frac {1}{z-\omega} \partial_{\omega}) {\hat \theta}_{u+v}^t(\omega) 
\,\,\,\,+ :\!{\hat \theta}_u^t(z){\hat \theta}_v^t(\omega)\!:
\end{equation}
\begin{equation}
\label{eq4.2e}
{\hat \theta}_u^G(z){\hat {\bar \psi}}_v(\omega) =  ( -\frac {1}{(z-\omega)^2} 
+ \frac {1}{z-\omega} \partial_{\omega}) {\hat {\bar 
\psi}}_{u+v}(\omega)\,\,\,\, + :\!{\hat \theta}_u^G(z){\hat {\bar 
\psi}}_v(\omega)\!:
\end{equation}
\begin{equation}
\label{eq4.2f}
{\hat \theta}_u^G(z){\hat \psi}_v(\omega) = ( \frac {2}{(z-\omega)^2} 
+ \frac {1}{z-\omega} \partial_{\omega}) {\hat 
\psi}_{u+v}(\omega)\,\,\,\, + :\!{\hat \theta}_u^G(z){\hat 
\psi}_v(\omega)\!:
\end{equation}
\begin{equation}
\label{eq4.2g}
{\hat J}_u^B(z){\hat \psi}_v(\omega) = \frac {1}{(z-\omega)^2} {\hat J}_{u+v}(\omega) 
+ \frac {1}{z-\omega} {\hat \theta}_{u+v}^t(\omega)\,\,\,\, + :\!{\hat 
J}_u^B(z){\hat \theta}_v^t(\omega)\!:
\end{equation}
\begin{eqnarray}
\label{eq4.2h}
{\hat J}_u^B(z){\hat J}_v^B(\omega) &= &\frac {20}{(z-\omega)^3} \sum_x \partial_{\omega} {\hat {\bar \psi}}_x(\omega) {\hat {\bar \psi}}_{u+v-x}(\omega) \nonumber \\
& &+ \frac {10}{(z-\omega)^2} \sum_x \partial_{\omega}^2 {\hat {\bar \psi}}_x(\omega) {\hat {\bar \psi}}_{u+v-x}(\omega) \nonumber \\
& &+ \frac {3}{z-\omega} \partial_{\omega} \left( \sum_x \partial_{\omega}^2 {\hat {\bar \psi}}_x(\omega){\bar \psi}_{u+v-x}(\omega)\right) \nonumber \\
& &+ :\!{\hat J}_u^B(z){\hat J}_v^B(\omega)\!: 
\end{eqnarray}
In Eqs. (4.1) and (4.2), the symbol $:\cdot:$ is operator-product 
normal ordering in the twisted sectors of the orbifolds. The operators
 $\{{\hat {\bar \psi}}_u$, ${\hat \psi}_u\}$ and $\{{\hat J^{B}}_{u}\}$ with 
 ${\bar u} = 0,1$ are respectively the 
 {\it twisted reparametrization ghost fields} and the {\it twisted 
 BRST currents}. The twisted reparametrization ghosts can also be 
 obtained by the Faddeev-Popov procedure [29] and choice of the 
 conformal gauge from the extended actions with
 $\mathbb{Z}_{2}$-twisted permutation gravity of Ref. [16]. The 
twisted operator products also record that the orbifold central charges 
 $\hat c= -52,52$ and $0$ of the various extended,twisted stress tensors 
 ${\hat\theta}_{u}^{G}(\mbox{ghost}), {\hat\theta}_{u}(\mbox{matter})$ and  
 ${\hat\theta}_{u}^{t}(\mbox{total})$ have not changed from those of 
 the corresponding eigen-stress tensors. The extended matter stress 
 tensors ${\hat \theta}_{u}$ in particular are the operator versions 
 of the classical (conformal-gauge) extended stress tensors of Ref. 
 [16], and 
 the explicit forms of these operators will be given for all $\hat c=52$ matter
 in the following paper of this series. 

In the orbifold program the mode formulation of each twisted sector 
is left to last, where the moding follows from the monodromies of the 
local fields. In this case, we find that the mode expansions  
\renewcommand{\theequation}{\thesection.\arabic{subsection}\alph{equation}}
\setcounter{subsection}{3}
\setcounter{equation}{0}
\begin{equation}
\label{eq4.3a}
{\hat {\bar \psi}}_u(z) = \sum_{m \in {\mathbb Z}} {\hat c}_u( m + \tfrac {u}{2}) z^{-( m + \frac {u}{2}) + 1},
\quad {\bar u} = 0,1
\end{equation}
\begin{equation}
\label{eq4.3b}
{\hat \psi}_u(z) = \sum_{m \in {\mathbb Z}} {\hat b}_u(m + \tfrac {u}{2}) z^{-( m + \frac {u}{2})-2}
\end{equation}
\begin{equation}
\label{eq4.3c}
{\hat \theta}_u^t = \sum_m {\hat L}_u^t( m + \tfrac {u}{2}) z^{-(m + 
\frac {u}{2})-2},\,\,\, {\hat L}_u^t( m + \tfrac {u}{2}) 
= {\hat L}_u( m + \tfrac {u}{2}) + {\hat L}_u^G( m + \tfrac {u}{2})
\end{equation}
\begin{equation}
\label{eq4.3d}
{\hat J}_u(z) = \sum_m {\hat J}_u( m + \tfrac {u}{2}) z^{-( m + \frac 
{u}{2})-1},\,\,\, {\hat J}_u^B(z) 
= \sum_m {\hat J}_u^B( m + \tfrac {u}{2}) z^{-(m + \frac {u}{2})-1}
\end{equation}
follow from the monodromies (4.1d), where $\{{\hat c}_{u}, {\hat 
b}_{u}\}$  and  $\{{\hat J^{B}}_{u}\}$ with $\bar 
u=0,1$ are respectively the 
twisted reparametrization-ghost modes and the 
twisted BRST current modes. Then the operator product system (4.2) and 
standard orbifold contour 
integrations (see e.g. Ref. [5]) give the mode algebras of the new 
twisted BRST systems:
\renewcommand{\theequation}{\thesection.\arabic{subsection}\alph{equation}}
\setcounter{subsection}{4}
\setcounter{equation}{0}
\begin{equation}
\label{eq4.4a}
[{\hat c}_u( m + \tfrac {u}{2}),{\hat b}_v( n + \tfrac {v}{2})]_+ = 
\delta_{m+n+\frac {u+v}{2},0}\quad\quad \bar u,\bar v \in \{0,1\}
\end{equation}
\begin{equation}
\label{eq4.4b}
[{\hat c}_{u}(m+\tfrac{u}{2}),{\hat c}_{v}(n+\tfrac{v}{2})]_{+} = [{\hat 
b}_{u}(m+\tfrac{u}{2}),{\hat b}_{v}(n+\tfrac{v}{2})]_{+} = 0
\end{equation}
\begin{eqnarray}
\label{eq4.4c}
[{\hat L}_u(m + \tfrac {u}{2}),{\hat L}_v( n + \tfrac {v}{2})] &=
 &( m - n + \tfrac {u-v}{2}){\hat L}_{u+v}( m+n+ \tfrac {u+v}{2})+ \nonumber \\
& &\hspace{-.2in}+ \tfrac{52}{12} ( m + \tfrac {u}{2})(( m + \tfrac {u}{2})^2 - 1) \delta_{m+n+\frac {u+v}{2},0} 
\end{eqnarray}
\begin{eqnarray}
\label{eq4.4d}
[{\hat L}_u^G( m + \tfrac {u}{2}),{\hat L}_v^G( n + \tfrac {v}{2})] &=
 &( m - n + \tfrac {u-v}{2}) {\hat L}_{u+v}^G( m + n + \tfrac {u+v}{2})+ \nonumber \\
& &\hspace{-.2in}-\tfrac{52}{12}( m + \tfrac {u}{2}) (( m + \tfrac {u}{2})^2 - 1) \delta_{m+n+\frac {u+v}{2},0}  
\end{eqnarray}
\begin{equation}
\label{eq4.4e}
[{\hat L}_u^t( m + \tfrac {u}{2}),{\hat L}_v^t( n + \tfrac {v}{2})] = ( m - n + \tfrac {u-v}{2}) {\hat L}_{u+v}^t( m + n + \tfrac {u+v}{2})
\end{equation}
\begin{equation}
\label{eq4.4f}
[{\hat L}_u^G( m + \tfrac {u}{2}),{\hat c}_v(n + \tfrac {v}{2})] = -(2( m + \tfrac {u}{2}) + n + \tfrac {v}{2}) {\hat c}_{u+v}( m + n + \tfrac {u+v}{2})
\end{equation}
\begin{equation}
\label{eq4.4g}
[{\hat L}_u^G( m + \tfrac {u}{2}),{\hat b}_v( n + \tfrac {v}{2})] = ( m - n + \tfrac {u-v}{2}){\hat b}_{u+v}( m + n + \tfrac {u+v}{2})
\end{equation}
\begin{equation}
\label{eq4.4h}
[{\hat J}_u^B( m + \tfrac {u}{2}),{\hat b}_v( n + \tfrac {v}{2})]_+ =
 ( m + \tfrac {u}{2}) {\hat J}_{u+v}( m + n + \tfrac {u+v}{2}) 
 + {\hat L}_{u+v}^t( m + n + \tfrac {u+v}{2})
\end{equation}
\begin{eqnarray}
\label{eq4.4i}
[ {\hat J}_u^B( m \!\!\!\!&+&\!\!\!\! \tfrac {u}{2}),{\hat J}_v^B( n + \tfrac {v}{2})]_+  =  
\nonumber \\
& = & \{ 10( m + \tfrac {u}{2})( n + \tfrac {v}{2}) 
- 3( m + n + \tfrac {u+v}{2})( m + n + \tfrac {u+v}{2} + 1)\}\times \nonumber \\
& \times &\!\! \sum_x \sum_{p \in {\mathbb Z}} ( p + \tfrac {x}{2})
 {\hat c}_x( p + \tfrac {x}{2}) {\hat c}_{u+v-x}( m + n - p + \tfrac {u+v-x}{2}). 
\end{eqnarray}
I call attention in particular to the three distinct order-two orbifold 
Virasoro algebras [1,27,9,12] in Eqs. (4.4c),(4.4d) and (4.4e), whose integral 
Virasoro subalgebras are generated by $\{{\hat L}_0(m)\}$, $\{{\hat 
L}_0^G(m)\}$ 
and $\{{\hat L}_0^t(m)\}$ with central charges $52$, $-52$ and $0$ 
respectively.

The new twisted BRST systems constructed here are a central 
result of this paper. The systems are complete for all the twisted 
open-strings at matter central charge $\hat c=52$ and, with the 
addition of a right-mover copy, all the twisted closed strings at 
$\hat c=52$ are described as well. The universal form of these systems reflects 
their common origin in the $\mathbb{Z}_{2}$-twisted permutation 
gravity [16] which governs all $\hat c=52$ strings. The extended matter 
Virasoro generators
$\{ {\hat L}_u( m + \frac {u}{2}),\ {\bar u} = 0,1\}$ of course vary
from sector to sector
of these orbifolds, but I 
defer the explicit form of these generators to the next paper of 
this series. 

Using these results, I turn next to some algebraic properties of the new BRST 
operator $\hat Q$ itself:
\renewcommand{\theequation}{\thesection.\arabic{subsection}\alph{equation}}
\setcounter{subsection}{5}
\setcounter{equation}{0}
\begin{equation}
\label{eq4.5a}
{\hat Q} \equiv {\hat J}_0^B(0)
\end{equation}
\begin{equation}
\label{eq4.5b}
{\hat Q}^2 = 0
\end{equation}
\begin{equation}
\label{eq4.5c}
[{\hat Q},{\hat b}_u(m + \tfrac {u}{2})]_+ = {\hat L}_u^t(m+\tfrac {u}{2})
\end{equation}
\begin{equation}
\label{eq4.5d}
[{\hat Q},{\hat L}_u^t(m + \tfrac {u}{2})] = 0,\quad {\bar u} = 0,1
\end{equation}
In particular,the nilpotency (4.5b) follows from Eq. (4.4i), the 
anticommutator (4.5c) follows from Eq. (4.4h), and the commutator 
(4.5d) is a consequence of (4.5b) and (4.5c) together. As noted above 
for the untwisted BRST system, other nilpotent modes of the twisted 
BRST current $\hat J^{B}$ are implied by the anticommutator (4.4i). 

Finally, we may use the new BRST operator and Eq. (4.5c) to define the {\it physical states} 
$\{|\chi\rangle\}$ in each twisted sector as follows:
\renewcommand{\theequation}{\thesection.\arabic{subsection}\alph{equation}}
\setcounter{subsection}{6}
\setcounter{equation}{0}
\begin{equation}
\label{eq4.6a}
{\hat Q}|\chi\rangle =\,\, {\hat b}_u((m+\tfrac {u}{2}) \ge 0) 
|\chi\rangle\,\, =\,\, {\hat c}_u((m+\tfrac {u}{2}) > 0) |\chi\rangle = 0
\end{equation}
\begin{equation}
\label{eq4.6b}
\to \quad {\hat L}_u^t((m+\tfrac {u}{2}) \ge 0) |\chi\rangle = 0,\quad {\bar u} = 0,1.
\end{equation}
The $\{\hat c\}$ condition in (4.6a) is not used here, but will play a 
role in the discussion below.

\setcounter{section}{4}
\section{The Extended Physical State Conditions}
\label{sec5}

To go further, we need more explicit forms of various operators in 
the system. For the twisted reparametrization ghosts, I introduce the 
following mode normal-ordered product
\renewcommand{\theequation}{\thesection.\arabic{equation}}
\setcounter{equation}{0}
\begin{eqnarray}
\label{eq5.1}
:{\hat A}_u(m+\tfrac {u}{2}){\hat B}_v(n+\tfrac {v}{2}):_M &\equiv &-\theta(m+\tfrac {u}{2} > 0){\hat B}_v(n + \tfrac {v}{2}){\hat A}_u(m + \tfrac {u}{2}) \nonumber \\
& &+ \tfrac {1}{2} \delta_{m + \tfrac {u}{2},0}\, [{\hat A}_0(0),{\hat B}_v(n+\tfrac {v}{2})] \nonumber \\
& &+ \theta(m+\tfrac {u}{2} < 0) {\hat A}_u(m + \tfrac {u}{2}) {\hat B}_v(n + \tfrac {v}{2}) 
\end{eqnarray}
where ${\hat A}$ and ${\hat B}$ can be either ${\hat c}$ or ${\hat b}$.

Then the mode expansions in Eq. (4.3) straightforwardly give (see 
e.g. Ref. [5]) the alternate form of the ${\hat {\bar \psi}},{\hat \psi}$
operator product:
\renewcommand{\theequation}{\thesection.\arabic{subsection}\alph{equation}}
\setcounter{subsection}{2}
\setcounter{equation}{0}
\begin{equation}
\label{eq5.2a}
{\hat {\bar \psi}}_u(z){\hat \psi}_v(\omega)\, =\, \delta_{u+v,0\!\!\!\!\mod 
2} \,{\hat \Delta}_{\bar u}(z,\omega)\,\, +\,\,\, :\!{\hat {\bar \psi}}_u(z){\hat 
\psi}_v(\omega)\!:_M
\end{equation}
\begin{eqnarray}
\label{eq5.2b}
{\hat \Delta}_{\bar u}(z,\omega) &= &(\frac {z}{\omega})^{\frac {\bar u}{2} + 1} \frac {1}{z-\omega} 
+ \frac {z}{2\omega^2} (1 - {\bar u}),\,\,\,\,\, {\bar u} = 0,1 \\
\label{eq5.2c}
&= &\frac {1}{z-\omega} + a_0(\omega) + (z-\omega)a_1(\omega,{\bar u}) + 
\mathcal{O}(z-\omega)^2
\end{eqnarray}
\begin{equation}
\label{eq5.2d}
a_0(\omega) = \frac {3}{2\omega},\quad a_1(\omega,{\bar u}) = \frac {4-{\bar u}}{8\omega^2}.
\end{equation}
Comparing this result with Eq. (4.2a), one obtains the following 
relations between the two kinds of normal-ordered products
\renewcommand{\theequation}{\thesection.\arabic{subsection}\alph{equation}}
\setcounter{subsection}{3}
\setcounter{equation}{0}
\begin{equation}
\label{eq5.3a}
:\!{\hat {\bar \psi}}_u(z){\hat \psi}_v(z)\!:\,\,\, =\,\,\, :\!{\hat {\bar 
\psi}}_u(z){\hat \psi}_v(z)\!:_M +\,\,\, \frac{3}{2z}\, \delta_{u+v,0\!\!\!\!\mod 2} 
\end{equation}
\begin{equation}
\label{eq5.3b}
:\!{\hat {\bar \psi}}_u(z) \partial_z {\hat \psi}_v(z)\!:\,\,\, =\,\,\, :\!{\hat {\bar 
\psi}}_u(z)\partial_z {\hat \psi}_v(z)\!:_M +\,\,\,\frac {{\bar 
u}-16}{8z^2}\, \delta_{u+v,0\!\!\!\!\mod 2}
\end{equation}
\begin{equation}
\label{eq5.3c}
:\!\partial_z{\hat {\bar \psi}}_u(z){\hat \psi}_v(z)\!:\,\,\, =\,\,\,
 :\!\partial_z {\hat {\bar \psi}}_u(z){\hat \psi}_v(z)\!:_M +\,\,\, \frac 
 {4-{\bar u}}{8z^2}\, \delta_{u+v,0\!\!\!\!\mod 2}
  \end{equation}
where I remind that $:\!\cdot\!:$ is operator-product normal 
ordering. These relations and Eq. (4.1a,b) give the mode-ordered form 
of the operators in the twisted ghost system
\renewcommand{\theequation}{\thesection.\arabic{subsection}\alph{equation}}
\setcounter{subsection}{4}
\setcounter{equation}{0}
\begin{equation}
\label{eq5.4a}
{\hat J}_u^G(z) \equiv \,\,{\hat J}_u(z) - \frac {3}{z} 
\delta_{u,0\!\!\!\!\mod 2}\,\, =\,\, \sum_{m \in {\mathbb Z}} J_u^G(m + \tfrac {u}{2}) z^{-(m+\frac {u}{2})-1}
\end{equation}
\begin{equation}
\label{eq5.4b}
{\hat J}_u^G(m+\tfrac {u}{2})\,\, =\,\, \sum_v \sum_{p \in {\mathbb Z}} 
:\!{\hat c}_v(p + \tfrac {v}{2}){\hat b}_{u-v}(m-p+\tfrac {u-v}{2})\!:_M
\end{equation}
\begin{eqnarray}
\label{eq5.4c}
{\hat J}_0^G(\!\!\!\!\!&m&\!\!\!\!=0) = \tfrac {1}{2} [{\hat c}_0(0),{\hat b}_0(0)] + \sum_{p=1}^{\infty} ({\hat c}_0(-p){\hat b}_0(p) 
- {\hat b}_0(-p){\hat c}_0(p)) + \nonumber \\
& &\hspace{.2in} - {\hat b}_{-1}(-\tfrac {1}{2}) {\hat c}_1(\tfrac 
{1}{2}) + \nonumber \\
 & &\!+ \sum_{p=1}^{\infty} ({\hat c}_1(-p+\tfrac {1}{2}) {\hat b}_{-1}(p-\tfrac {1}{2}) 
- {\hat b}_{-1}(-p-\tfrac {1}{2}){\hat c}_1(p + \tfrac {1}{2})) 
\end{eqnarray}
\begin{equation}
\label{eq5.4d}
{\hat \theta}_u^G(z) = \sum_v :\!{\hat {\bar \psi}}_v(z) \partial_z {\hat \psi}_{u-v}(z) 
+ 2\partial_z {\hat {\bar \psi}}(z) {\hat \psi}_{u-v}(z)\!:_M - \,\,\frac 
{17}{8z^2}\,\, \delta_{u,0\!\!\!\!\mod 2}
\end{equation}
\begin{eqnarray}
\label{eq5.4e}
\!\!{\hat L}_u^G(m + \tfrac {u}{2}) \!&= &\!\!-\sum_v \sum_{p \in {\mathbb Z}} 
(m+p+\tfrac {u+v}{2}) :\!{\hat c}_v(p + \tfrac {v}{2}) {\hat 
b}_{u-v}(m-p+ \tfrac {u-v}{2})\!:_M \nonumber \\
& &\hspace{.2in}- \tfrac {17}{8} \delta_{m + \tfrac {u}{2},0} 
\end{eqnarray}
where $\{{\hat J}_u^G(z)\}$ is the properly-ordered form of the twisted 
ghost current. The $\{{\hat c}_{0},{\hat b}_{0}\}$ terms  of the ghost charge ${\hat J}_0^G(0)$
in Eq. (5.4c) are isomorphic to the untwisted ghost charge in Eq. 
(2.5c). Using Eqs. (4.1c), (4.2e) (4.3) and (5.1), mode-ordered expressions 
can also be written out for the twisted BRST current $\{{\hat J}_u^B\}$ 
and the BRST charge $\hat Q$, but these will not be needed in the 
present development.

The mode-ordered forms in Eq. (5.4) and the ghost-mode conditions on 
the physical states in 
Eq. (4.6a) then imply the further characterization of the physical states
\renewcommand{\theequation}{\thesection.\arabic{subsection}\alph{equation}}
\setcounter{subsection}{5}
\setcounter{equation}{0}
\begin{equation}
\label{eq5.5a}
({\hat J}_u^G((m+\tfrac {u}{2}) \ge 0) + \tfrac {1}{2} \delta_{m + \tfrac {u}{2},0}) |\chi\rangle = 0
\end{equation}
\begin{equation}
\label{eq5.5b}
({\hat L}_u^G((m + \tfrac {u}{2}) \ge 0) + \tfrac {17}{8} \delta_{m + \tfrac 
{u}{2},0}) |\chi\rangle = 0,\quad {\bar u} = 0,1
\end{equation}
in terms of the twisted ghost current and the extended Virasoro 
generators of the twisted ghost system.

Finally, Eqs. (4.6b) and (5.5b) give the desired description of the 
physical states of each twisted sector in terms of the extended 
Virasoro generators of the $\hat c=52$ matter:
\renewcommand{\theequation}{\thesection.\arabic{equation}}
\setcounter{equation}{5}
\begin{equation}
\label{eq5.6}
({\hat L}_u((m + \tfrac {u}{2}) \ge 0) - \tfrac {17}{8} \delta_{m + \tfrac 
{u}{2},0}) |\chi\rangle = 0,\quad {\bar u} = 0,1.
\end{equation}
Here and in the succeeding papers of this series, I refer to this 
result as the {\it extended physical-state condition} (or 
extended Virasoro condition) for $\hat c=52$ string matter. The two 
components $\bar u=0,1$ of this condition are the operator analogues of
the extended classical Virasoro constraints $\{{\hat \theta}_u = 0$, 
${\bar u} = 0,1\}$
implied by the corresponding extended actions with $\mathbb{Z}_{2}$-twisted permutation
gravity in Ref. [16].

\setcounter{section}{5}
\section{Discussion}
\label{sec6}

This paper begins the discussion of the orbifold-strings of 
permutation-type at the operator level. In particular, I have 
constructed here the twisted BRST systems and extended physical-state 
conditions of all orbifold-string matter at central charge $\hat 
c=52$, which corresponds to the special case of $\mathbb{Z}_{2}$-twisted permutation 
gravity in the extended actions of Ref. [16].

I remind the reader that each of the twisted $\hat c=52$ strings has 
twice the conventional number of negative-norm states, and that the 
two-component form of the extended physical-state condition (5.6) supports
our speculation that the orbifold-strings of permutation-type can be
free of negative-norm states at higher central charge. The explicit form of the extended 
matter Virasoro generators and the physical spectrum of the $\hat c=52$
strings will be discussed in the following paper of the series. In 
later papers, we will also see that the extended physical-state 
condition follows as expected from extended Ward identities in 
the interacting theories.

The derivation given here used the standard operator techniques of the 
orbifold program [1-15], leaving for another time the more-involved Faddeev-Popov
 derivation of the twisted BRST systems from 
the extended actions of Ref. [16]. Using the operator-product form of 
untwisted BRST in Sec. 2, the standard operator techniques can be 
straightforwardly applied to obtain the twisted BRST systems of 
the general orbifold of permuta-tion-type -- whose twisted sectors couple
to general permutation-twisted world-sheet gravity [16]. I have not yet 
worked these systems out but, as an educated guess, I would expect the 
following labelling of the twisted BRST currents $\{\hat J^{B}\}$ and BRST
charges $\{\hat Q\}$ in twisted open-string sector $\sigma$ of the general orbifold of 
permutation-type
\renewcommand{\theequation}{\thesection.\arabic{subsection}\alph{equation}}
\setcounter{subsection}{1}
\setcounter{equation}{0}
\begin{equation}
\label{eq6.1a}
{\hat J}_{{\hat \jmath}j}^B(m + \tfrac {\hat 
\jmath}{f_{j}(\sigma)}),\quad {\hat 
Q}_j \equiv {\hat J}_{0j}^{B}(0)
\end{equation}
\begin{equation}
\label{eq6.1b}
[{\hat Q}_i,{\hat Q}_j]_+ = 0
\end{equation}
\begin{equation}
\label{eq6.1c}
{\hat c} = 26 K,\quad {\bar {\hat \jmath}} = 
0,1,\dots,f_j(\sigma)-1,\quad \sum_j  f_j(\sigma) = K
\end{equation}
as well as a right-mover barred copy for the twisted closed-string sectors.
In this labelling $f_j(\sigma)$ is the length of cycle $j$ in the corresponding 
element $h_{\sigma} \in H(\mbox{perm})$, which permutes $K$ copies of 
the closed string $U(1)^{26}$ in the untwisted sector. In particular Eq. (6.1)
shows a total of $K$ 
twisted BRST currents in each twisted sector $\sigma$, including a BRST charge for each cycle in 
$h_{\sigma}$. This reduces for $K=2$ to the case studied in this 
paper because the non-trivial element of $\mathbb{Z}_{2}$ is a 
single cycle of length $2$.

\section*{Acknowledgements}

For helpful information,  discussions and encouragement, I thank L. 
Alvarez-Gaum$\acute{e}$, K. Bardakci, I. Brunner,
J. de Boer, D. Fairlie, O. Ganor, E. Gimon, C. Helfgott, E. Kiritsis, R. 
Littlejohn, S. Mandelstam, J. McGreevy, N. Obers, A. Petkou, E. 
Rabinovici, V. Schomerus, K. Schoutens, C. Schweigert and E. Witten. This work was 
supported in part by the Director, Office of Energy Research, Office of 
High Energy and Nuclear Physics, Division of High Energy Physics of the U.S.
Department of Energy under Contract DE-AC02-O5CH11231 and in part by the National
Science Foundation under grant PHY00-98840.

\end{document}